# Data center top of rack switch to multiple spine switches optical wireless uplinks

Abrar S. Alhazmi[1], Osama Zwaid Alsulami[1], T. E. H. El-Gorashi[1],
Mohammed T. Alresheedi[2] and Jaafar M. H. Elmirghani[1]

[1]School of Electronic and Electrical Engineering, University of Leeds, LS2 9JT, United Kingdom
[2] Department of Electrical Engineering, King Saud University, Riyadh, Kingdom of Saudi Arabia
elasal@leeds.ac.uk, ml15ozma@leeds.ac.uk , t.e.h.elgorashi@leeds.ac.uk ,
malresheedi@ksu.edu.saj m.h.elmirghani@leeds.ac.uk

**ABSTRACT**
Infrared (IR) uplinks can achieve high data rates, which are essential in a range of applications. This paper introduces a novel approach to enable data centre uplink communication. We introduce a novel method to enable communication between racks and spine switches. In our proposed data centre, we consider three racks, each of which has its own angle diversity transmitter (ADT) that is located on top of the rack. Four wide field of view receivers are fixed to the ceiling of the data centre. Each such receiver is connected to a spine switch. We evaluate the performance of our proposed system when each link operates at a data rate above 2.8 Gb/s. Multiple links can be used to achieve higher data rates using the space or wavelength dimensions. The results show that our proposed system has the ability to work at a high data rate with good performance while using simple on-off-keying.

**Keywords**: Angle diversity receivers; Data centers; Uplink design; Optical wireless

## 1. INTRODUCTION

Optical wireless communication (OWC) systems were proposed more than 30 years ago as a type of broadband technology [1]. Wireless information transmission through OWC links uses the optical spectrum to unlock higher information rates, which can be used for six-generation (6G) communication frameworks. OWC is a promising solution that can permit much higher information rates in 6G. OWC systems have gained the interest of many researchers [2], [3]. In OWC, infrared (IR) or visible light (VL) can be used to send the data from the transmitter to the receiver [4].

OWC systems use laser diodes, or light-emitting diodes, to transfer light from a transmitter through free space to an optical receiver and finally to fiber optic systems [5], [6]. More benefits can be obtained using IR compared with radio frequency (RF) [7], [8]. These advantages include immunity against interference caused by nearby electrical devices and a wider transmission bandwidth. In addition, IR systems have shorter transmission ranges compared with RF systems [9]. This makes IR systems ideal for indoor environments. Moreover, IR systems offer a higher data rate compared with RF systems, making them suitable for systems that intrinsically require high data rates, such as data centers. Many studies have demonstrated that OWC frameworks can transmit data at rates reaching 25 Gb/s in indoor environments [10] – [21]. Nevertheless, OWC have only recently been considered as candidates for commercial applications when the data requirements increased and the cost of both the transmitter and the receiver components became affordable [22] – [24].

It should be noted that in data center uplinks, transmitters are located on top of the racks, whereas receivers are placed on the ceiling. Uplink OWC systems were studied in [23], [24]. Spot diffusing schemes can be used to improve the signal to noise ratio [25] – [29]. In addition, the performance of the system can be more improved by using diversity approaches [30] – [33].

In this paper, we propose and design an angle diversity transmitter (ADT) to achieve high data rate links for uplink communication in the data center. In the data center, we considered three racks, each of which had its own ADT. In addition, four wide field of view (WFOV) receivers were installed on the ceiling of the data center linking to the four spine switches.

The remainder of this paper is organized as follows: Section 2 presents the data center structure and Section 3 describes the optical receiver design. Section 4 presents the optical transmitter structure. Section 5 discusses the simulation environment and results. The paper's conclusions are given in Section 6.

## 2. DATA CENTER STRUCTURE

In our simulation, the data center is divided into pods and each pod has dimensions 8 m × 8 m × 3 m (length × width × height), which are similar to the pod dimensions in [34], as shown in Figure 1. As the figure illustrates, the simulation uses a vacant pod without any windows or doors, and the pod is assumed to be unfurnished. In this work, the data center comprises three racks located at (1 m, 1 m, and 0.25 m), (4 m, 4 m, and 0.25 m), and (4 m, 7 m, and 0.25 m). The rack dimensions are 0.6 m x 1.2 m x 1.75 m, and each rack has a top-of-rack (ToR) switch. In addition, the data center has four equally spaced spine switches fixed to the room's ceiling. An ADT is located on top of each rack.

An ADT is located at the top of each rack, directing its line-of-sight beams to the receivers. Figure 1 illustrates the data center with multiple racks, and each rack has a ToR switch. There are multiple spine switches linking to the receivers in the ceiling. The racks are raised 0.25 m above the main floor.

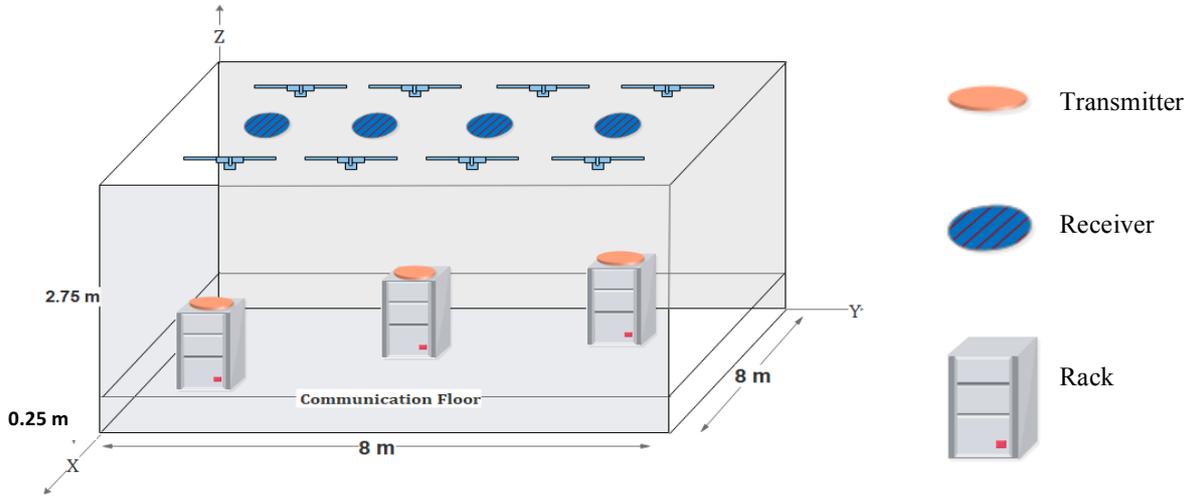

**Figure 1.** The data center setup.

## 3. OPTICAL RECEIVER DESIGN

In this paper, a row of optical receivers is located along the middle of the ceiling (see Figure 1) at four equally spaced locations, namely, (4, 1, 3), (4, 3, 3), (4, 5, 3), and (4, 7, 3). We used a WFOV receiver in the present work, which has a 90° field of view (FOV), with an area equal to 20 mm$^2$ to support high data rates. In addition, the responsivity of the optical receiver was 0.6 A/W [35], [36].

## 4. OPTICAL TRANSMITTER DESIGN

In this study, we employed an ADT, as used in [37] – [39]. The ADT consisted of four branches. Two angles were used to assign the direction of each branch of the ADT, namely, azimuth and elevation. In our ADT, the azimuth angles of the branches of the ADT at the corner of the data center (1.3 m, 1.6 m, 2 m), are set to 348°, 27°, 51°, and 63°; and the elevation angles are set at 20°, 18°, 13°, and 9°. The azimuth angles of the second ADT at the center of the data center (4 m, 4 m, 2 m) are set to 270°, 270°, 90°, and 90°; and the elevation angles are set to 18°, 45°, 45°, and 18°. The azimuth angles of the third ADT transmitter at the data center pod corner, ie at (4 m, 6.7 m, 2 m) are set to 270°, 270°, 270°, and 90°, and elevation angles sare et to 10°, 15°, 30°, and 73°. In addition, the semi-angle at half power of each branch is set to 2°. It should be noted that the azimuth, elevation, and semi-angle at half power of the ADT were selected to enable each branch of the ADT to serve one receiver. Each branch of each ADT emits 150 mW [40], [41] optical power.

## 5. SIMULATION SETUP AND RESULTS

In this section, we evaluate the performance of our proposed ADT for the data center links. The simulation was carried out using ray tracing [42]. The results are presented in terms of the signal-to-noise ratio (SNR) and data rate. Every individual receiver was served by one branch of an ADT.

Observing the rack locations in Figure 1, a rack at the center of the data center has a very short link to the spine switch just above it but has longer links to the spine switches at the edges of the room. Similarly, the racks at the edges of the room have short links to the spine switches just above them but have very long links to the spine switches at the other edge of the room.

The SNR should be high in order to enable high data rates and to reduce the bit error rate (BER). For on-off keying (OOK), the BER is given as [43]:

$$\text{BER} = Q\sqrt{SNR} \tag{1}$$

The SNR assuming OOK modulation was computed using [44]:

$$SNR = \frac{(R(P_{s1}-P_{s0})^2)}{\sigma_t^2} \tag{2}$$

where $R$ is the photodetector responsivity ($R = 0.6 A/W$), $P_{s1}$ is the power associated with logic 1, $P_{s0}$ is the power associated with logic 0, and $\sigma_t^2$ is the total noise due to the received signal and receiver noise current spectral density. The receiver input noise current was 4.47 pA/√Hz for the 5 GHz receiver used [45]. The calculation of $\sigma_t^2$ can be found in [46], [47]:

$$\sigma_t^2 = \sigma_{pr}^2 + \sigma_{bn}^2 + \sigma_{sig}^2 \tag{3}$$

where $\sigma_{pr}^2$ is the mean square receiver preamplifier noise, $\sigma_{bn}^2$ is the mean square background shot noise and $\sigma_{sig}^2$ is the mean signal induced shot noise. Figure 2a provides a comparison of the SNR (dB) of the three ADT links, quoted as the SNR at each receiver (R1, R2, R3, and R4). As can be seen, our proposed data center links can provide a good data rate with a acceptable SNR. The SNR of each receiver (R1, R2, R3 and R4) when the links operate at a data rate of at least 2.8 Gb/s, is illustrated in Figure 2a. As can be seen, our proposed data center links can provide SNR not less than 15.6 dB, which is needed to achieve a BER of $10^{-9}$. Receiver one (R1) has the lowest performance. This is due to the fact that the distance between this receiver and its ADT is high. In addition, R2 and R3 have the same performance for transmitter 2, and R1 and R4 have the same performance for transmitter 2. It should be noted that a shorter communication distance leads to reduced power loss and, as a result, a higher SNR, while the links operate at a high data rate. Figure 2b shows the achievable data rate of our proposed uplink communication data center system, which is 7 Gb/s for all receiver from different racks except receiver 1 from rack 3 and receiver 4 from rack 1 which support around 2.8 Gb/s due to the long distance between the rack and receiver. This data rate can be increased by utilizing higher transmit powers for these links as the data centre pods are enclosed and eye safety can be handled by restricting access and by switching off the lasers when the doors open to allow human access.

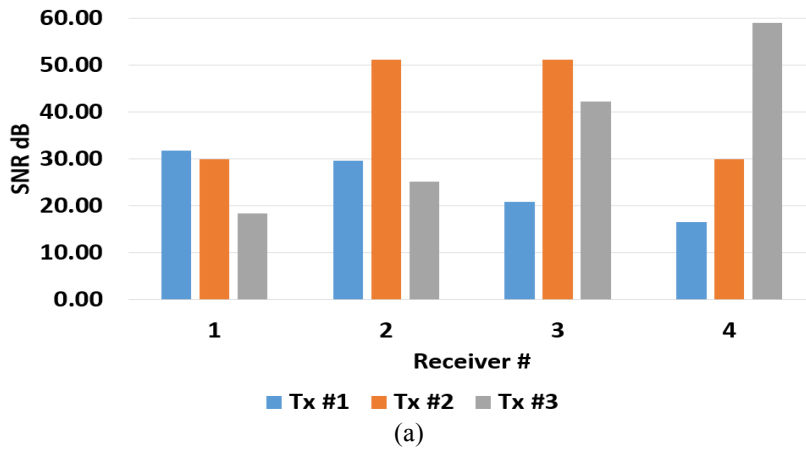

(a)

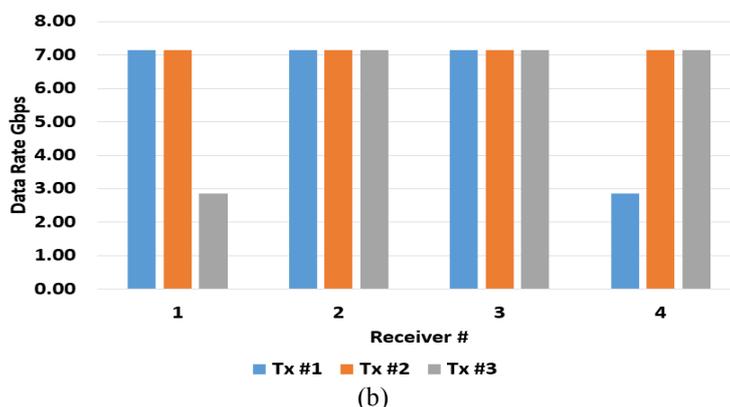
(b)

**Figure 2.** The optical wireless data centre system. (**a**) signal-to-noise ratio (SNR) and (**b**) data rate of angle diversity transmitter (ADT) with wide field of view receiver (WFOVR).

## 6. CONCLUSIONS

In this paper, we proposed, designed, and evaluated an ADT for high-data-rate data center links. The data center setup considered had three racks, and each rack has its own ADT located at the top of the rack. Four receivers were located on the ceiling linking to spine switches, and each receiver was served by one branch of the ADT. The results showed that our proposed data center links worked at a data rate of 7 Gb/s while some receivers achieved just 2.8 Gb/s, with SNR not less than 15.6 dB.

**ACKNOWLEDGEMENTS**

The authors would like to acknowledge funding from the Engineering and Physical Sciences Research Council (EPSRC) INTERNET (EP/H040536/1), STAR (EP/K016873/1) and TOWS (EP/S016570/1) projects. The authors extend their appreciation to the deanship of Scientific Research under the International Scientific Partnership Program ISPP at King Saud University, Kingdom of Saudi Arabia for funding this research work through ISPP#0093. ASA would like to thank Taibah University in the Kingdom of Saudi Arabia for funding her PhD scholarship. OZA would like to thank Umm Al-Qura University in the Kingdom of Saudi Arabia for funding his PhD scholarship, all data are provided in full in the results section of this paper.